\documentclass[twocolumn,epsfig,graphics,showpacs,floatfix,mathbbm]{revtex4}

\usepackage{amsmath,amsfonts,amssymb,graphics,graphicx,epsfig,color,times,bbm}
\usepackage{amsthm}
\usepackage{psfrag}
\graphicspath{{figures/}}

\newcommand{\me}{\mathrm{e}}
\newcommand{\mi}{\mathrm{i}}
\newcommand{\md}{\mathrm{d}}
\renewcommand{\vec}[1]{\boldsymbol{#1}}

\newcommand{\cc}{\mathbbm{C}}

\newcommand{\id}{\mathbbm{1}}

\newcommand{\tr}{\operatorname{tr}}

\begin{document}
\bibliographystyle{apsrev}

\title{Exact relaxation in a class of non-equilibrium quantum lattice systems}

\author{M.\
Cramer$^{1,2}$,
C.M.\ Dawson$^{1,2}$,
J.\
Eisert$^{1,2}$,
T.J.\ Osborne$^{3}$}

\address{1 Blackett Laboratory, Imperial College London, Prince
Consort Road, London SW7 2BW, UK,\\
2 Institute for
Mathematical Sciences, Imperial College London, Prince's
Gardens, London
SW7 2PE, UK,\\
3 Department of Mathematics, Royal Holloway University of
London, Egham, Surrey TW20 0EX, UK}

\date{\today}

\begin{abstract}
A reasonable physical intuition in the study of interacting quantum
systems says that, independent of the initial state, the system will
tend to equilibrate. In this work we study a setting where
relaxation to a steady state is exact, namely for the Bose-Hubbard
model where the system is quenched from a Mott quantum phase to the strong superfluid regime. We find that the evolving state
\emph{locally} relaxes to a steady state with maximum entropy
constrained by second moments, maximizing the
entanglement, to a state which is different from the thermal state
of the new Hamiltonian. Remarkably, in the infinite system limit
this relaxation is true for all large times, and no time average is
necessary. For large but finite system size we give a time interval
for which the system locally  ``looks relaxed" up to a prescribed
error. Our argument includes a central limit theorem for harmonic
systems and exploits the finite speed of sound. Additionally, we
show that for all periodic initial configurations, reminiscent of
charge density waves, the system relaxes locally. We sketch
experimentally accessible signatures in optical lattices as well as
implications for the foundations of quantum statistical mechanics.
\end{abstract}

\maketitle

\section{Introduction}
\label{sec:intro} The study of the non-equilibrium properties of
quantum many body systems has recently entered a renaissance era.
This has been motivated, in part, by recent experimental
developments; the rapid progress of experiments involving ultracold
atoms in optical lattices, with their high degree of control and
long coherence times, has opened the door to precise experimental
studies of the dynamics of strongly interacting quantum systems
\cite{Experiments}. In addition, questions of relaxation and
thermalisation for non-equilibrium systems are again receiving
attention from foundational perspectives. This is partly triggered
by intuition from quantum information theory where maximally or
almost maximally entangled states emerge from appropriate
distributions of random states
[2--4].

One particularly fascinating setting which has recently received
intensive study is that of \emph{quenching}, that is, a sudden
change of interaction strength
[5--12].
Several explanations for, and numerical studies of, quenched
dynamics have gradually led to the formulation of a rather general
body of theory and conjectures; it has been mooted that if the
system starts in the ground state of one Hamiltonian then certain
properties such as correlators of the system should relax to an
analogue of the thermal state of the new Hamiltonian after 
a quench
[5, 7--9].
Thus we are motivated by these observations to formulate a
conjecture. This \emph{local relaxation conjecture} states 
that a system should locally relax to a steady state, respecting 
conserved quantities of motion.

The local relaxation conjecture may sound suspiciously like a
violation of unitarity as the global state must, of course, remain
pure throughout the dynamics. However, a reasonable physical
intuition which explains why there is no contradiction is that for a
small block of sites the rest of a system acts like a reservoir and
thus allows the site to maximise its entropy subject to the
constraint that the energy is preserved. Indeed, the full
explanation for the emergence of a local steady state during the
course of the quench may be intuitively described along these lines:
as time evolves the system becomes correlated \cite{Entanglement} --
from each site a wavefront moving at the speed of sound for the
lattice emerges carrying information. As time progresses more and
more excitations will have passed through a given site, see Fig.\ 1.
The cumulative effect of these successive excitations results in an
effective averaging process; information stored in one site becomes
infinitely diluted across the lattice as time progresses. A similar
intuition has been recently emphasised by Calabrese and Cardy
\cite{calabrese} in the context of quenching to a critical system.

\begin{figure}
  \includegraphics[width=10cm]{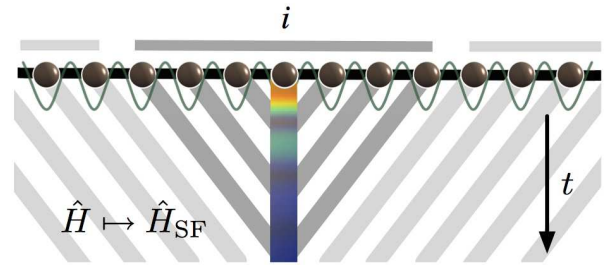}
  \caption{\label{fig:intuition}Intuitive picture of the relaxation process in the quenched
  Bose-Hubbard model as discussed in the main text: For
  any lattice site $i$ (or any block of sites) within a
  cone (dark grey) defined by the speed of
  sound excitations
  significantly contribute to the local mixing of the state at
   the site. Contributions from
  outside this cone (light grey)
  are exponentially suppressed.
  The incommensurate influence of the lattice sites in the cone
  gives rise to a relaxation to maximal entropy for large times $t$.
  When the excitations have travelled through the entire lattice,
  recurrences will occur.}
\end{figure}

In this work we introduce a physical setting where the local
relaxation conjecture can be studied analytically. We find an exact
convergence of all local states to a steady state for long times in
the quenched Bose-Hubbard model. We imagine the system starts in a
Mott insulator phase and is then suddenly switched to the deep
superfluid regime. Our method is self-contained and physically
motivated, and is valid also for finite system sizes relevant to experimental settings. In the course of the proof we are also able to quantify the dilution of information
throughout the lattice. Thus, our approach is considerably
simpler and more physically intuitive than a potential approach
based on the $C^*$-algebraic arguments developed in
Ref.\  \cite{Robinsonandfriends} to study the local relaxation
of free fermions and bosons freely moving in $\mathbb{R}^n$
and classical systems \cite{Spohn}.
Interestingly, the convergence 
is not only true in the {\it time average}, but
actually true for any large instant of time. This is in
contrast to the recent approaches developed in quantum information
\cite{Brodyandco}, where, in order to study this problem, it would
seem necessary to consider the time-averaged local state rather than
the local state itself. Additionally, our approach does not require
the system being quenched to to be critical --- we obtain the same
results regardless of whether the system being quenched to is
critical or not. For blocks of sites we also find a relaxation, but
not to the thermal state of the new Hamiltonian.

\section{Quenched Bose-Hubbard dynamics}
\label{sec:bhm}
We start from a
Bose-Hubbard Hamiltonian,
modeling, for example, a dilute gas of ultra-cold atoms which are
Bose-condensed in an optical lattice \cite{Classic}.
In one dimension (generalizations of our findings to higher dimensions are entirely straightforward) it reads $\hat{H}=\hat{H}_{\text{SF}}+\hat{U}$, where
\begin{equation}
    \hat{H}_{\text{SF}}=-J \sum_{\langle i, j\rangle} \hat{b}_i^\dagger \hat{b}_j,
    \;
    \hat{U}=\frac{U}{2} \sum_{i=1}^{N}
    \hat{n}_i(\hat{n}_i-1)  - \mu
    \sum_{i=1}^{N}\hat{n}_i.
\end{equation}
Here $\hat{b}_1,\dots, \hat{b}_{N}$ denote the bosonic annihilation operators,
and $\hat{n}_i = \hat{b}_i^\dagger \hat{b}_i$. The coefficient $J$ is the {\it
hopping
matrix element}, governing the strength of hopping to neighboring
sites ($\langle i, j\rangle$ indicates summation over nearest
neighbors) and $U$ defines the strength of the on-site interaction. Finally, the chemical potential $\mu$ controls the particle number. For theoretical simplicity we shall assume that that the system is translationally invariant with periodic boundary conditions, so that the underlying lattice is a ring.
This Hamiltonian  exhibits two distinct
phases as $J$ and $U$ are varied.
When the hopping dominates, that is $J \gg U$, the system is in a
{\it superfluid phase}. For a dominant on-site repulsion the ground state
is a {\it Mott insulator}.

We imagine that the system is initially held at chemical potential
$\mu$ and is in the ground state in the Mott regime \cite{Classic}.
In this deep Mott phase (corresponding to $J=0$) the ground state
vector is a product $ |m\rangle^{\otimes N}$ of individual number
state vectors $|m\rangle$ of $m$ bosons at each site. Here, $m=\max\{0, [ (\mu/U+1)/2 ] \}$, where $[\cdot]$ denotes the closest integer to the value in brackets. We
then imagine that the system is rapidly {\it quenched}, at $t=0$, to
the strong superfluid regime $J \gg U$. We model this quench by an
instantaneous change in the Hamiltonian to the regime $U = 0$. Thus
the system state vector at time $t$, $|\psi(N,t)\rangle$,   is given by (we set $\hbar=1$)
\begin{equation}
    |\psi(N,t)\rangle = \me^{-\mi t \hat{H}_{\text{SF}}} | m \rangle^{\otimes N}.
\end{equation}


\section{Relaxation for a single site}
\label{sec:single_site}
As outlined earlier, we are interested in whether subsystems
consisting of single sites or blocks thereof equilibrate, and, if
so, in what sense. For clarity
we start by proving relaxation for the quantum
state $\hat{\varrho}_i(N,t)$ of a single site $i$ in the lattice $L=[1,\dots,N]$, and then
extend this result to multiple sites and different initial conditions.
The state of site $i$ is given by a partial trace
\begin{eqnarray}
    \hat{\varrho}_i(N,t)  = \tr_{L\setminus\{i\}} \bigl[|\psi(N,t)\rangle\langle\psi(N,t)|\bigr].
\end{eqnarray}
Throughout this paper we will describe the system in phase
space, making use of the characteristic function to represent the state $\hat{\varrho}_i$. It is defined as
\begin{equation}\label{eq:charfun}
    \chi_i(\alpha;N,t) = \tr[
    \hat{\varrho}_i(N,t) e^{\alpha \hat{b}_i^\dagger - \alpha^*\hat{b}_i}
    ],
\end{equation}
where $\alpha\in \mathbb{C}$.
We now turn to showing that we indeed locally, {\it at each site} $i$,
find convergence to a
state that maximises the entropy: We prove that the state
$\hat{\varrho}_i(N,t)$ approximates a Gaussian state $\hat{\varrho}_G$ arbitrarily well. We will find
that  the first moments of $\hat{\varrho}_G$
vanish,
$\tr[\hat{\varrho}_G\hat{b}_i]=0$,
and its second moments are all conserved
and identical to the
initial ones,
\begin{eqnarray}
    \tr[\hat{\varrho}_G\hat{b}_i\hat{b}_i]&=
    &\langle m|\hat{b}_i\hat{b}_i|m\rangle=0,\\
    \tr[\hat{\varrho}_G\hat{n}_i]&=&\langle m|\hat{n}_i|m\rangle=m.
\end{eqnarray}
Gaussian states
maximise the local entropy for fixed second moments, which are
constants of motion for the initial states $|m\rangle^{\otimes N}$ under consideration. For more general initial states exhibiting time dependent second moments see Section~\ref{sec:moreGeneral}.

Our main result can thus be summarised as
\begin{equation}
    \hat{\varrho}_i(N,t) \rightarrow  \hat{\varrho}_G,
\end{equation}
where the limit of large $N$ is taken first and then
the limit of large $t$ is taken --- note that there is no time average involved. More precisely, for any $\varepsilon>0$
and any recurrence time $t_{\text{Recurrence}}>0$
we can find a system size $N$ such that
\begin{equation}
\label{eq:togaussian}
     \| \hat{\varrho}_i(N,t) - \hat{\varrho}_G\|_{\text{tr}}<\varepsilon
     \text{ for } t\in [t_{\text{Relax}},t_{\text{Relax}}+t_{\text{Recurrence}}].
\end{equation}
Here, the relaxation time $t_{\text{Relax}}$ is
governed by the hopping strength $J$, which
defines  the speed of sound of the system, see Eq.~(\ref{eq:trange}) and Fig.~\ref{fig:intuition}. For finite $N$ recurrences occur for times larger than $t_{\text{Recurrence}}$, however these
can be shifted to infinity for large $N$.
Closeness of $\hat{\varrho}_i(N,t)$ and $\hat{\varrho}_G$ is measured in trace norm which of course
means that \emph{all} local expectation
values are the same as for the relaxed state. Another way of interpreting the relaxation to a Gaussian state
is that the entanglement between the site $i$ and the rest of the chain becomes maximal.

For clarity, we deliberately discuss this case of a single lattice
site first in great detail. We will show Eq.~(\ref{eq:togaussian})
by evaluating the corresponding limits for the characteristic
function $\chi_i$, and showing that it tends to a
Gaussian in $\alpha$ for large times $t$ and lattice sizes $N$.
Pointwise convergence in the characteristic function has been shown
to imply trace norm convergence of the corresponding density
operators \cite{Cushen,Trace}, so this is sufficient to prove
Eq.~(\ref{eq:togaussian}). There are four main steps in the
approach: (i) We take and expand the logarithm of the characterisic
function, then (ii) we identify certain terms quadratic in $\alpha$.
(iii) The magnitude of the remaining terms are then bounded, and
finally (iv) we make use of these bounds to  evaluate the limiting
behavior of the characteristic function. Readers interested in the
final result may safely skip this part of the argument, which is
completed by Eq.~(\ref{eq:chigaussian}).

Employing the cyclic rule of trace, we find
\begin{eqnarray}\nonumber
    \chi_i(\alpha;N,t) &=&\tr\left[\me^{-\mi t\hat{H}_{\text{SF}}}|m
    \rangle^{\otimes N}\langle  m |^{\otimes N}
    \me^{\mi t\hat{H}_{\text{SF}}}
    e^{\alpha \hat{b}_i^\dagger - \alpha^*\hat{b}_i} \right]\\
    &=&\langle  m |^{\otimes N}\me^{\mi t\hat{H}_{\text{SF}}}
    e^{\alpha \hat{b}_i^\dagger - \alpha^*\hat{b}_i} \me^{-\mi t\hat{H}_    {\text{SF}}}|m\rangle^{\otimes N}.\label{eq:chi_2}
\end{eqnarray}
Now, using the Baker-Hausdorff identity (or, alternatively, by solving Heisenberg's equation of motion) one finds for the Heisenberg
representation of $\hat{b}_i$
\begin{equation}
    \label{eq:bdynamics}
    \me^{\mi t\hat{H}_{\text{SF}}}\hat{b}_i
    \me^{-\mi t\hat{H}_{\text{SF}}}=
    \sum_{j=1}^{N}\left[V   (N,t)\right]_{i,j}
    \hat{b}_j,\; V(N,t)=\me^{-\mi t \mathcal{J}(N)},
\end{equation}
where the $N\times N$ circulant
matrix $\mathcal{J}(N)$ represents
the hopping operator $\hat{H}_{\text{SF}}$.
Its entries read
$\mathcal{J}_{i,j}=-J\delta_{\text{dist}(i,j),1}$, where we
denote by $\text{dist}(i,j)$ the distance between two sites $i$ and $j$. Due to this circulant structure $\mathcal{J}(N)$ may be diagonalized via a discrete Fourier transform to give eigenvalues
\begin{equation}
    \lambda_k =-2J\cos\left(2\pi k/N\right),
\end{equation}
and thus
\begin{equation}
\label{eq:formV}
\left[V(N,t)\right]_{i,j}=\frac{1}{N}\sum_{k=1}^{N}\me^{-\mi t \lambda_k}\me^{2\pi\mi k(i-j)/N}.
\end{equation}
The characteristic function in Eq.~(\ref{eq:chi_2}) is then given by
\begin{equation}
    \chi_i(\alpha;N,t) =
    \prod_{j=1}^{N}  \langle m |\me^{\alpha [V(N,t)]_{i,j}^*
    \hat{b}^\dagger_j -
    \alpha^*[V(N,t)]_{i,j} \hat{b}_j}| m \rangle.
\end{equation}
A straightforward calculation (see, for example, Ref.\
\cite{Barnett97a}) shows that the
expectations appearing on the right-hand side
evaluate to Laguerre polynomials
and we may write
\begin{equation}
\label{eq:chi_laguerre}
    \chi_i(\alpha;N,t) =  \me^{-|\alpha|^2/2} \prod_{j=1}^{N} L_m( |\beta_{i,j}(N,t)  |^2),
\end{equation}
with $\beta_{i,j}(N,t) = \alpha [V(N,t)]^*_{i,j}$ (we will often assume implicit dependence on $t$ and $N$ and
simply write $\beta_{i,j}$ and $V_{i,j}$).

The matrix elements $V_{i,j}$ encode all the information about the
dynamics. We identify each row of this matrix as the quantum state
of a free particle initially localised at site $i$ which is hopping
on the ring. As time progresses the initially localised particle
disperses rapidly throughout the ring. In deriving the bounds of
step (iii) it will be essential that we can identify for any
$\epsilon>0$ those times $t$ and lattice sizes $N$ for which we can
bound $|V_{i,j}| <\epsilon$. This part has, in turn, two main
ingredients: We make use of (iiia) a
{\it ``central limit type argument''}
as well as of (iiib) a {\it  ``Lieb-Robinson type argument''}.
In this way,
we also have a very clear handle on the case of finite $N$, which is
the situation encountered in any experiment. These two regimes of
(iiia) and (iiib) correspond to lattice sites close and far away
from the distinguished site $i$: For sites labeled $j$ with
$\text{dist}(i,j)\leq L(t)$ for some $L(t)$ we are able to grasp the
dynamics using a central limit argument. These are the sites
actually contributing to the relaxation process (see Fig.\ 1). Sites
with $\text{dist}(i,j)>L(t)$ also contribute to the local dynamics at site
$i$, but this contribution is exponentially suppressed due to the
finite speed of sound for the system.

We start by investigating the sites with negligible contribution,
case (iiib). As $\mathcal{J}_{i,j}=0$ for $\text{dist}(i,j)>1$, we have that $[\mathcal{J}^n]_{i,j}=0$ for $\text{dist}(i,j)>n$ and thus
\begin{equation}
\begin{split}
    V_{i,j}
    &= \left[e^{-i\mathcal{J}t}\right]_{i,j}
    =\sum_{n=0}^\infty \frac{(-it)^n}{n!}\left[\mathcal{J}^n\right]_{i,j} \\
    &=\sum_{n\ge \text{dist}(i,j)} \frac{(-it)^n}{n!}\left[\mathcal{J}^n\right]_{i,j}.
\end{split}
\end{equation}
Now, for any matrix $M$ one has $|M_{i,j}|\le \| M\|$, where $\|\cdot\|$ indicates the operator norm, and $\|\mathcal{J}\|=2J$. Hence,
\begin{eqnarray}\nonumber
    \left|V_{i,j}\right| & \leq &
    \sum_{n\ge \text{dist}(i,j)} \frac{(2Jt)^n}{n!}\leq
    \sum_{n\ge \text{dist}(i,j)} \left(\frac{6Jt}{n}\right)^n\nonumber
    \\
    &\leq&
    \sum_{n\ge \text{dist}(i,j)} \left(\frac{6Jt}{\text{dist}(i,j)}\right)^n,
\end{eqnarray}
where we used the fact that $n!\ge(n/3)^n$.
We thus finally find that contributions from sites $j$ with $\text{dist}(i,j)=d>6Jt$
are exponentially suppressed:
\begin{equation}
\left|V_{i,j}\right|\leq\frac{(6Jt/d)^d}{1-6Jt/d},
\end{equation}
independent of the system size $N$. This is the intuition provided
by what are known as {\it Lieb-Robinson bounds}: Sites beyond the
cone defined by the speed of sound will not significantly alter the
state at site $i$ (see again Fig.~\ref{fig:intuition}). From this
expression we see that, if we require $|V_{i,j}| < \epsilon$, then
it is sufficient that $\text{dist}(i,j)=d > L(t)$, where $L(t)$ is given by
the solution to $(6Jt/L(t))^{L(t)}=\epsilon(1-6Jt/L(t))$. A 
crude bound on $L(t)$ may be obtained by noting that
\begin{equation}
\frac{(6Jt/d)^d}{1-6Jt/d}\le\frac{6Jt/d}{1-6Jt/d},
\end{equation}
i.e., for given $\epsilon >0$ we have
\begin{equation}
\label{eq:boundhi}
|V_{i,j}| < \epsilon \text{ for all } i,j:\text{dist}(i,j)>L(t)=6Jt\frac{1+\epsilon}{\epsilon}.
\end{equation}

We now turn to the case where $ \text{dist}(i,j)\leq L(t)$, case (iiia). Writing $V_{i,j} = V_l$ for $l = i-j$, we recall that
\begin{equation}
\label{eq:Vjk}
    V_{l} =  \frac{1}{N} \sum_{k=1}^{N}
    \me^{2 \mi t J\cos (2\pi k/N)} \me^{2\pi \mi k l /N}.
\end{equation}
In the limit $N\rightarrow\infty$ this approaches an integral representation of the Bessel function
\begin{equation}
J_l(x)=\frac{1}{2\pi\mi^l}\int_0^{2\pi}\!\!\!\!\md\phi\;\me^{\mi x\cos(\phi)}\me^{\mi l\phi},
\end{equation}
up to a phase $\mi^l$. For finite $N$, the $V_{l}$ can thus be
thought of as Riemann sum approximations to $\mi^lJ_l$. Our strategy
for choosing finite $N,t$ such that $|V_{l}| < \epsilon$ will make
use of a bound on the error involved in such an approximation,
together with the bound $|J_l(x)| < x^{-1/3}$ for all $x\geq0 $ and
all $l$ \cite{Landau00a}. Consider the quantity $|V_{l} - i^l
J_l(2Jt)|$. On using the Riemann sum approximation, we find
\begin{equation}
    \left|V_{l} - i^l J_l(2Jt)\right| \leq 2 \pi^2 (2Jt + l)/ N ,
\end{equation}
and by combining this with the bound on $|J_l(2Jt)|$ we obtain
\begin{equation}
\label{eq:boundlow}
    \left|V_{l}\right|
    \leq \frac{2 \pi^2}{N} (2Jt + L(t)) + (2Jt)^{-1/3},
\end{equation}
complementing the bound in Eq.~(\ref{eq:boundhi}).

\begin{figure}
  \includegraphics[width=7cm]{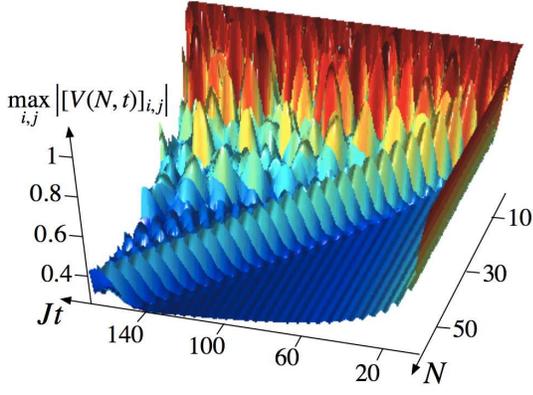}
  \caption{\label{fig:V}The entries of the matrix governing the dynamics of the system are arbitrarily small---the system is locally arbitrarily close to a Gaussian state---for sufficiently large $N$ and a time interval $[t_{\text{Relax}},t_{\text{Relax}}+t_{\text{Recurrence}}]$ (compare Eq.~(\ref{eq:trange}) and Fig.~\ref{fig:intuition}). After excitations have traveled through the whole lattice recurrences occur at $t=t_{\text{Recurrence}}$, which is given by the speed of sound. For $N\rightarrow\infty$ no recurrences occur, $t_{\text{Recurrence}}\rightarrow\infty$, and locally the system relaxes to a Gaussian for large times.}
\end{figure}

Combining both bounds we find that {\em for all} $i,j$ (see
Fig.~\ref{fig:V}, where we plot $\max_{i,j}|V_{i,j}|$ versus $N$ and
$t$)
\begin{equation}
\label{eq:trange}
\left|V_{i,j}\right| < \epsilon \text{ for all } t:
    \frac{4}{\epsilon^3} < Jt < \frac{N}{4\pi^2 }\frac{\epsilon^2}{8\epsilon+6}.
\end{equation}
For any $\epsilon > 0$, this provides sufficient conditions for
satisfying $|V_{i,j}|<\epsilon$ for all $i,j$ and finite $N$ and
$t$. This ends step (iii), the intuition of which is summarized in
Fig.~\ref{fig:intuition}.

We now turn to our original aim, that of showing that the
characteristic function $\chi_i$ tends to a Gaussian in
$\alpha$. The steps in the proof follow the basic structure of the
classical proof of the central limit theorem \cite{Williamsbook}. We
start by taking the logarithm of the right hand side of Eq.\
(\ref{eq:chi_laguerre}). Expanding the logarithm, we then make use
of the observations on $|V_{i,j}|$ obtained above, together with
certain properties of Laguerre polynomials.

We can always find appropriate $t$ and $N$ such that
$0<L_m(|\beta_{i,j}| ^2)\le 1$ as $|\beta_{i,j}|$ can be made
arbitrarily small for given $\alpha$ and all $i,j$, see
Eq.~(\ref{eq:trange}). Thus we may write
\begin{eqnarray}
\log \prod_{j=1}^{N} L_m(|\beta_{i,j}| ^2)
= -\sum_{j=1}^{N} \sum_{k=1}^\infty \frac{\left(1- L_m(|\beta_{i,j}|^2)\right)^k}{k}. \label{eq:log_expansion}
\end{eqnarray}
The Laguerre polynomials are defined as
\begin{equation}
L_m(x) = \sum_{n=0}^m {m \choose m-n} \frac{(-x)^n}{n!}
=:1 - mx + l_m(x),
\end{equation}
where we defined $l_m(x)$ for later use.
Hence
\begin{eqnarray}
\log  \prod_{j=1}^{N} L_m(|\beta_{i,j}|^2)&=&
 - m \sum_{j=1}^{N} |\beta_{i,j}|^2+ \sum_{j=1}^{N} l_m(|\beta_{i,j}|^2)\nonumber\\
\label{eq:log_expansion}
 &-&
     \sum_{j=1}^{N}\sum_{k=2}^\infty \frac{\left(1- L_m(|\beta_{i,j}|^2)\right)^k}{k}.
\end{eqnarray}
From the definition of $\beta_{i,j}$ and the
unitarity of $V$, the first summation evaluates to $ - m |\alpha|^2$,
which is quadratic as desired, and the remainder of the proof
consists of showing that the remaining terms tend to zero for given $\alpha$.

Since $|\beta_{i,j}| = | \alpha V_{i,j}|$,  we may use the results above to ensure that for any fixed $\alpha$ the $|\beta_{i,j}|$ are as small as desired, and in particular we shall first suppose that $|\beta_{i,j}| < 1$. In this case we have $|l_m(|\beta_{i,j}|^2)| \leq |\beta_{i,j}|^4 l_m(-1)$,  and it follows that
\begin{equation}
\label{eq:sum_lm_bound}
\begin{split}
    \left |\sum_{j=1}^{N} l_m(|\beta_{i,j}|^2) \right |
    &\leq l_m(-1)|\alpha|^4\sum_{j=1}^{N} \left|V_{i,j}\right|^4\\
    &\leq l_m(-1)|\alpha|^4 \sup_{i,j} \left|V_{i,j}\right |^2\\
&    =: l_m(-1)|\alpha|^4\bar{V}.
    \end{split}
\end{equation}
The large $N$ behavior of this term will evidently be determined by the corresponding behavior of $\bar{V}$, for which we already know that it tends to zero if we first let $N$ and then $t$ go to infinity, see Eq.~(\ref{eq:trange}).

Now we study the last term in Eq.~(\ref{eq:log_expansion}):
\begin{equation}
    \begin{split}
    \label{eq:sum_fm_bound}
    &\hspace{-1cm}\left|-\sum_{k=2}^{\infty}
    \frac{(1-L_m(|\beta_{i,j}|^2))^k}{k}\right|\\
    &=(1-L_m(|\beta_{i,j}|^2)\sum_{k=1}^{\infty}
    \frac{(1-L_m(|\beta_{i,j}|^2))^k}{k+1}\\
    &\leq -\left(1 - L_m(|\beta_{i,j}|^2)\right) \log L_m(|\beta_{i,j}|^2).
    \end{split}
\end{equation}
By again expanding the Laguerre polynomial as above we have
\begin{eqnarray}
    -\log L_m(|\beta_{i,j}|^2)\nonumber
    &\le&
    -\log\left(1-m|\beta_{i,j}|^2-|l_m(|\beta_{i,j}|^2)|\right)\nonumber\\
    &\le&-
    \log\left(1-|\alpha|^2\bar{V}(m+ l_m(-1))\right) \nonumber\\
    & =:& \Lambda,
\end{eqnarray}
which is a sensible bound if we require that $|\alpha|^2\bar{V}(m+
l_m(-1)) < 1$, which can again be ensured with appropriate $N$ and
$t$. Furthermore, $\bar{V}\rightarrow 0$, i.e., also
$\Lambda\rightarrow 0$. We thus finally obtain the upper bound
\begin{eqnarray}
\nonumber
\Lambda
\sum_{j=1}^{N}\left(1 - L_m(|\beta_{i,j}|^2)\right)&
=\Lambda
\sum_{j=1}^{N}\left(m|\beta_{i,j}|^2 - l_m(|\beta_{i,j}|^2)\right)\\
&\le \Lambda\left(m|\alpha|^2+|\alpha|^4\bar{V}l_m(-1)\right)
\end{eqnarray}
for the absolute value of the last term in 
Eq.~(\ref{eq:log_expansion}).

Equipped with these bounds,
we arrive at the desired statement: We find for any
$\alpha\in \mathbb{C}$,
\begin{equation}
\label{eq:chigaussian}
    \chi_i(\alpha;N,t) = \me^{- \left(m + 1/2\right)|\alpha|^2}
    + f(\alpha;N,t),
\end{equation}
where $|f(\alpha;N,t)|$ can be made arbitrarily small for
all $t\in [t_{\text{Relax}},t_{\text{Recurrence}}+t_{\text{Relax}}]$
and a suitable choice of the lattice size $N$, see Eq.~(\ref{eq:trange}).
Since pointwise closeness for the characteristic functions of two quantum
states translates to closeness in trace norm for the
states themselves \cite{Cushen,Trace}, 
we have completed the argument for a single site.

We therefore find that the state of a single site, in the limit of
large $t$, maximises the entropy for the given initial 
second moments: Gaussian states have this extremal property
of having maximal entropy. Another way of saying this is that, 
subject to this constraint, it is {\it maximally
entangled} with the rest of the chain. Since we can 
bound the time scale on which relaxation
occurs -- the single site relaxes to a Gaussian state at the rate
$(2Jt)^{-1/3}$ --  our argument is robust under small
perturbations.

This result also gives a clear handle on the situation for finite
system sizes: For any finite system size $N$, there will of course
eventually be recurrences. The time scale for such recurrences to
occur in the $(N,t)$ plane is governed by the speed of sound and the
system size $N$: when $t < cN$, where $c$ is the speed of sound,
there will be no recurrences, and our argument will apply, giving
rise to the time $t_{\text{Recurrence}}$, see Eq.~(\ref{eq:trange}).
Thus we also arrive at a precise way to describe relaxation for {\it
infinite system size}: locally the system ``looks exactly
relaxed'' for large times, apparently giving the impression of an
 ``irreversible dynamics''. The same result also holds for
correlation functions. This {\it local relaxation} is
perfectly consistent with the unitary dynamics of the full system;
the whole system never truly relaxes to an equilibrium state as it
must always preserve the {\it full memory} of the initial condition.

The Gaussian character of the state can also be viewed as resulting
from a different type of central limit theorem (cf. Ref.\
\cite{Wolf}). The intuition is that time evolution can be thought of
as a sequence of of beam splitters acting in parallel, i.e.\ a
quantum quincunx or Galton's board (this follows from the fact that
the dynamics are well-approximated by a quantum cellular automaton).
After the bosons bounce through the quincunx we should arrive at a
Gaussian characteristic function in phase space describing the state
of any single site.


\section{Relaxation of blocks}
\label{sec:blocks}
The equilibration for multiple sites, which includes blocks of
lattice sites, may be shown using identical techniques, and here we
briefly describe the necessary, but minor, modifications. The
characteristic function as a function of $\vec{\alpha}=(\alpha_1, \ldots,
\alpha_{s})^T\in\cc^s$ for a set of sites $S = [1, \dots, s]$ with
reduced state
$\hat{\varrho}_S$ is defined as
\begin{equation}
\label{eq:def_mm_chi} \chi_S(\vec{\alpha})
 = \tr\left[\hat{\varrho}_S(N,t) \me^{\sum_{i=1}^{s}
 (\alpha_i \hat{b}_i^\dagger - \alpha_i^* \hat{b}_i}) \right] .
\end{equation}
After the quench the time-evolving bosonic operators are given by
Eq.~(\ref{eq:bdynamics}), and the characteristic function becomes
 \begin{equation}
        \chi_S(\vec{\alpha};N,t) =
    \me^{-\vec{\alpha}^\dagger\vec{\alpha}/2 }
    \prod_{j=1}^{N} L_m(|\beta_{S,j}(N,t)|^2),
\end{equation}
where the $\beta_{S,j}$ are now given by $\sum_{i=1}^{s} \alpha_i V_{i,j}(N,t)$. Note that otherwise Eq.~(\ref{eq:chi_laguerre}) is unchanged, and the argument can be followed as before to show pointwise convergence in $\vec{\alpha}$. Using arguments essentially identical to the ones in the proof for a single site, we find in the same sense
a convergence
\begin{equation}
    \hat{\varrho}_S(N,t)\rightarrow  \hat{\varrho}_G^{\otimes s}
\end{equation}
for large times $t$ and large system sizes $N$.
That is, the block tends to a product of maximal entropy states.

\section{Comparison with thermal states}

\label{sec:thermal_comparison} Let us contrast the state arising
from our dynamical relaxation with the thermal Gibbs state
$\me^{-(\hat{H}_{\text{SF}}-\mu\hat{N})/T}/\tr[  \me^{-(\hat{H}_{\text{SF}}-\mu\hat{N})/T} ] $ (setting $k_B=1$)
of the
new Hamiltonian $\hat{H}_{\text{SF}}$: for a thermal state at temperature $T>0$, 
we find the correlations \cite{Marcus}
\begin{equation}
   \langle \hat{b}_i^\dagger \hat{b}_j\rangle
   =  \left[ (\me^{(\mathcal{J}-\mu\id)/T}-\id )^{-1}\right]_{i,j}.
\end{equation}
The thermal state of $\hat{H}_{\text{SF}}$ is \emph{always} correlated for all
temperatures $T>0$ and all chemical potentials $\mu$ fixing the
particle number. In other words, although the system does mix and we
do arrive at a state with local maximal entropy, it does {\it not}
relax to the thermal state of the new Hamiltonian: the conserved
second moments prevent this from happening. The relaxation should be
thought of as resulting from a mixing process rather than of a
physical thermalisation process. Of course the second moments are
locally preserved by the dynamics, so under this constraint the
entropy is indeed maximised.

\section{More general initial states}
\label{sec:moreGeneral} 
\begin{figure}
  \includegraphics[width=6.8cm]{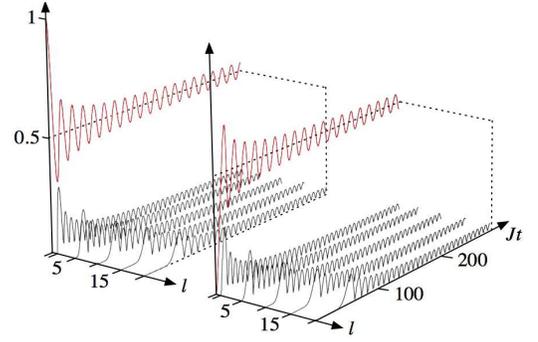}
  \caption{\label{fig:secondmoments}Second moments $\langle\hat{b}_1^\dagger\hat{b}_{1+l}\rangle$ (left) and $\langle\hat{b}_2^\dagger\hat{b}_{2+l}\rangle$ (right), $l=0,1,5,10,15,20$, for an initial
  periodic
  configuration $|0,1,0,1,\cdots ,0,1\rangle$
  on a lattice of size $N=300$.
  For larger times recurrences occur (not shown), while for $N\rightarrow\infty$, one has $\lim_{t\rightarrow\infty}\langle\hat{b}_i^\dagger\hat{b}_{j}\rangle=\delta_{i,j}/2$.}
\end{figure}

In fact, an argument along the lines of the
above proof can be carried out whenever the the initial state has
sufficiently rapidly decaying correlations, reminiscent of the
situation in the classical case \cite{Spohn}. Details of this
argument will be presented elsewhere \cite{Preparation}. For the
purposes of this article, we will now consider more general initial
states of the form $\otimes_{i=1}^{N} |m_i\rangle$, allowing for
different particle numbers at different sites -- allowing, e.g., for
configurations occurring in harmonic traps and
periodic distributions
with periodicity different from one (realized, e.g., by
superimposing a second lattice with different wavelength),
reminding of {\it charge density wave} or checkerboard
phases. For these inhomogeneous
initial states the second moments are no longer preserved, but for
certain initial states we can nevertheless prove a convergence of
second moments. While convergence of second moments is necessary for
a local relaxation, it is certainly not sufficient. However, as
before, we show that local states relax to a Gaussian state -- we
show that the characteristic function of the state tends to a
Gaussian in $\vec{\alpha}$ as higher cumulants can be shown to
vanish just like in the homogeneous case. For
initial states $|\psi(N,0)\rangle=\otimes_{i=1}^{N} |m_i\rangle$,
the expression for the characteristic function is
\begin{equation}
\label{eq:chi_laguerre_inhomog} \chi_S(\vec{\alpha};N,t) =
\me^{-\vec{\alpha}^\dagger\vec{\alpha}/2 }\prod_{j=1}^{N}
L_{m_j}(|\beta_{S,j}(N,t)|^2),
\end{equation}
with $|\beta_{S,j}|$ as above. Eqs~(\ref{eq:sum_lm_bound}-\ref{eq:sum_fm_bound}) and the surrounding discussion must then be modified to incorporate the different $m_j$, for example in the quantities $l_m(-1)$. All such quantities can be appropriately modified by taking their supremum over all the different $m_j$. We then find pointwise convergence of the characteristic function to
\begin{eqnarray}
    \nonumber
    &&\lim_{t\rightarrow\infty}
    \lim_{N\rightarrow\infty}\chi_S(\vec{\alpha};N,t)
    =\me^{-\vec{\alpha}^\dagger\vec{\alpha}/2}\\
    &&\hspace{1.0cm}
    \times\exp\Bigl(  - \lim_{t\rightarrow\infty} \lim_{N   \rightarrow\infty} \sum_{j=1}^{N} m_j |\beta_{S,j}|^2 \Bigr) ,
\end{eqnarray}
showing that there can be initial configurations for which the state never relaxes. For initial states with $m_j=m$ for almost all $j$, we always
find relaxation to a steady state as we also do for $P$-periodic initial configurations: Consider an initial state with $m_i=m_{i-P}$, completely determined by $m_1,\dots,m_P$.
Then, for a block $S$ of $s$ sites,
\begin{equation}
\sum_{j=1}^{N} m_j |\beta_{S,j}|^2=\sum_{k,l=1}^s\alpha_k\alpha_l^*
\sum_{p=1}^Pm_p\sum_{j=0}^{N/P-1}V_{k,Pj+p}^*V_{l,Pj+p}.
\end{equation}
Now,
\begin{eqnarray}
\sum_{j=0}^{N/P-1}V_{k,Pj+p}^*V_{l,Pj+p} &=&
\frac{1}{N^2}\sum_{r,s=1}^N
\me^{\mi t(\lambda_r-\lambda_s)}\me^{2\pi\mi (rl-sk)/N}\nonumber
\\
&\times&
\sum_{j=0}^{N/P-1}\me^{2\pi\mi Pj(r-s)/N},
\end{eqnarray}
and the last terms evaluate to a comb of Kronecker deltas,
\begin{equation}
    \sum_{j=0}^{N/P-1}\me^{2\pi\mi Pj(r-s)/N}=
    \frac{N}{P}\sum_{z=-P+1}^{P-1}\delta_{r-s,zN/P},
\end{equation}
which yields
\begin{eqnarray}
    &&\sum_{j=0}^{N/P-1} V_{k,Pj+p}^*V_{l,Pj+p}=
    \frac{\delta_{k,l}}{P}+
    \frac{1}{P}\sum_{\substack{z=-P+1\\ z\ne 0}}^{P-1}
    \me^{\pi\mi(\frac{z}{P}(l+k)+\frac{l-k}{4})}\nonumber\\
    &&\hspace*{1cm}\times
    \frac{1}{N}\sum_{r=1}^N\me^{-2\mi t\sin(\pi z/P)
    \lambda_r}\me^{2\pi\mi r(l-k)/N}.
\end{eqnarray}
Identifying the last line with $[V(N,-2t\sin(\pi z/P))]_{l,k}$, which tends to zero as $t$ and $N$ become large (see the proof for the relaxation of a single site), we find also in this scenario that the characteristic function tends to a Gaussian:
\begin{equation}
\lim_{t\rightarrow\infty}
\lim_{N\rightarrow\infty}\chi_S(\alpha;N,t)=\me^{-(1/2+\bar{m})\vec{\alpha}^\dagger\vec{\alpha}},\;\;
\hat{\varrho}_S\rightarrow\hat{\varrho}_G^{\otimes s},
\end{equation}
where $\bar{m}=\sum_{p=1}^Pm_p/P$. While initially different sites are uncorrelated, correlations build up over time and finally go to zero for large times  (see Fig.~\ref{fig:secondmoments}) and
the local state relaxes to a direct product of the same Gaussian at each site. 

\section{Integrable spin models and fermions}
\label{sec:fermions}
To further illustrate the generality of our approach, we sketch the
situation in other integrable models. We consider a $XY$ spin
chain,
\begin{equation}
\hat{H}=-\frac{1}{2}\sum_{\langle i,j\rangle}\left(\frac{1+\gamma}{4}\hat{\sigma}_i^x\hat{\sigma}_j^x+\frac{1-\gamma}{4}\hat{\sigma}_i^y\hat{\sigma}_j^y\right)-\frac{\lambda}{2}\sum_{i=1}^N\hat{\sigma}_i^z,
\end{equation}
again in the setting of a quenched interaction. By virtue of
the familiar Jordan-Wigner transformation, the Hamiltonian is
equivalent to a system of spinless fermions:
\begin{equation}
    \hat{H} =\frac{1}{2}\sum_{i,j}\left(\hat{f}_i^\dagger A_{i,j}\hat{f}_j
    -\hat{f}_i A_{i,j}\hat{f}_j^\dagger
    +\hat{f}_i B_{i,j}\hat{f}_j
    -\hat{f}_i^\dagger B_{i,j}\hat{f}_j^\dagger \right),
\end{equation}
where $A_{i,i}=\lambda$, $A_{i,j}=-1/2$ for $\text{dist}(i,j)=1$, $B_{i,j}=-B_{j,i}=\gamma/2$ for $\text{dist}(i,j)=1$ and zero otherwise.

We start with the system initially in one of the states $|m\rangle^{\otimes N}$, $m=0,1$, corresponding to the magnetized spin states $|\!\!\!\downarrow
\rangle^{\otimes N}$, $|\!\!\!\uparrow
\rangle^{\otimes N}$, and study time evolution
under the above Hamiltonian. As the initial states are fermionic Gaussian states, the system stays Gaussian for all times $t$. Hence,
we only need to consider seconds moments, which are most conveniently written in terms of Majorana operators, $\hat{m}_i=(\hat{f}_i+\hat{f}_i^\dagger)/\sqrt{2}$, $\hat{m}_{N+i}=\mi(\hat{f}_i^\dagger-\hat{f}_i)/\sqrt{2}$. Then
one may collect second moments in a correlation matrix $\gamma (N,t)$ with entries $[\gamma (N,t)]_{i,j}=\text{tr}[\hat{\varrho}(N,t)\hat{m}_i\hat{m}_j]$.
Time evolution of the initial state $|m\rangle^{\otimes N}$ under the above Hamiltonian then amounts to
\begin{equation}
    \gamma(N,t) = \me^{t H} \gamma(N,0)
    \me^{-t H},\, H=  \left[\begin{array}{cc}
    0  & V\\
    -V^T & 0
    \end{array}\right],
\end{equation}
where $V=A+B$ with eigenvalues
\begin{equation}
\lambda_k(V)=\lambda-\cos(2\pi k/N)-\mi\gamma\sin(2\pi k/N).
\end{equation}
A tedious but elementary calculation
delivers the resulting second moments
\begin{equation}
    \gamma(N,t) =\frac{1}{2}\left[\id +\mi (-1)^m\left(\begin{array}{cc}
    M_1(N,t)&M_2(N,t)\\
    -M_2^T(N,t)&M_1^T
    \end{array}\right)\right],
\end{equation}
where
\begin{eqnarray}
    M_1(N,t)&=&\sin(2|V|t)\frac{B}{|V|},\\
    M_2(N,t)&=&\id+\sin^2(|V|t)\left(\frac{V^2}{|V|^2}-\id\right),
\end{eqnarray}
from which we immediately see that for the isotropic $XY$ model
($\gamma=0$) second moments are a constant of motion,
$\gamma(N,t)=\gamma(N,0)$. We assume $\gamma\ne 0$ from now on.
Obviously, as in the bosonic case, the total state stays pure for
all times and its entropy is thus always zero, however, depending on
the system parameters (it will turn out that it depends on whether
the system being quenched to is critical or not -- in contrast to
the bosonic case, where relaxation occurs independent of the system
being critical or not), we find exact relaxation of second moments. To this
end, consider the second moments of the state of a single site,
$\hat{\varrho}_i(N,t)=\text{tr}_{L\setminus
\{i\}}[\hat{\varrho}(N,t)]$. It is completely determined by the
reduction of the correlation matrix to site $i$, i.e., we need to
consider $[M_1(N,t)]_{i,i}$ and $[M_2(N,t)]_{i,i}$. As the matrices
$A$ and $B$ commute and $M_1(N,t)$ is the product of a symmetric and
an antisymmetric matrix, we immediately find
$[M_1(N,t)]_{i,i}=0=[M_1(N,0)]_{i,i}$. For the sake of clarity, we
now focus on the Ising model, $\gamma=1$. It should be clear how the
following arguments may be
extended to explore the full parameter
space. We find
\begin{equation}
\begin{split}
\left[M_2(N,t)-\id\right]_{i,i}&=
\frac{1}{N}\sum_{k=1}^N\sin^2(|\lambda_k|t)\left(\frac{\lambda_k^2}{|\lambda_k|^2}-1\right)\\
&=\frac{1}{N}\sum_{k=1}^N\sin^2(|\lambda_k|t)\frac{\me^{4\pi\mi k/N}-1}{1-\lambda\me^{2\pi\mi k/N}}.
\end{split}
\end{equation}

Let us illustrate this for one particular non-critical case, namely
$\lambda=0$, for which one has $\lambda_k(V)=-\me^{2\pi\mi k/N}$ and
hence
\begin{equation}
\left[M_2(N,t)\right]_{i,i}=1-\sin^2(t),
\end{equation}
which never relaxes.

It turns out that for the non-critical case, $|\lambda|\ne 1$, the
system never locally relaxes. This is, of course, no surprise: in
the case where $\lambda = 0$ the Hamiltonian we quench to contains
only commuting terms. Physically, one of three different processes
can take place. In the first case, $\lambda < 1$, the system is
quenched to a state in the same phase. In this case we do not expect
local relaxation as the density of quasiparticles of, eg.,
$|\!\!\downarrow\rangle^{\otimes m}$ with respect to the new Hamiltonian
is too low: there are not enough quasiparticles to lead to an
averaging effect. In the critical case $\lambda = 1$, the
quasiparticles are delocalised and we expect that the density of
quasiparticles in $|\!\!\downarrow\rangle^{\otimes m}$ is high enough to
lead to a local relaxation. Finally, in the case $\lambda > 1$, the
quasiparticles of the new system are strongly localised, and appear
as pairs in the state $|\!\!\downarrow\rangle^{\otimes m}$ which then
oscillate locally.

Let us now turn to the critical case $\lambda=1$, where we expect relaxation. We find
\begin{eqnarray}
    &&\left[M_2(N,t)-\id\right]_{i,i}=-\frac{1}{N}
    \sum_{k=1}^N
    \sin^2(2t\sin(\pi k/N)) ,\nonumber\\
    &-&\frac{1}{N}
    \sum_{k=1}^N\sin^2(2t\sin(\pi k/N))
    \me^{2\pi\mi k/N},
\end{eqnarray}
where the first term approaches $-1/2$ and the second term goes to zero for large $N$, $t$, i.e., indeed, at each site the system relaxes when evolved in time under this critical Hamiltonian.
\section{Summary and discussion}
\label{sec:conclusion} In this work, we have investigated the
relaxation of Bose-Hubbard type systems and related models following
a sudden quench. The intuition we developed was that within a causal
cone of neighboring lattice sites, the incommensurate influence of
the propagating quasiparticles will give rise to a relaxation
dynamics. We have considered a setting in which we could prove an
exact local relaxation to a maximal entropy state, providing a
guideline of what is expected to happen in similar cases. Also, a
precise understanding follows for the scaling of this effect in the
system size, as well as of relaxation times. Indeed, the system
``looks'' perfectly relaxed, while at the same time keeping a
perfect memory of the initial situation. We hence can clarify the
difference between an apparent -- which does happen -- and an actual
global -- which does not happen (!) -- relaxation.
The argument presented here is expected to be valid whenever
the initial condition is sufficiently clustering, so if one has an
appropriate decay of correlations.

From the perspective of kinematical approaches to quantum
statistical problems, one can interpret our results as follows: When
randomizing over all possible pure states \cite{Winter}, the local
state will have maximal entropy for large systems. Here, we arrive
at the same result, but not by a {\it kinematical} argument, but via
a {\it dynamical} argument, where the mixing is achieved through
local physical dynamics in a lattice system. It would be interesting
to further try to combine the kinematical and dynamical pictures for
more general Hamiltonians.

Settings similar to those discussed here are readily accessible in
{\it experiments} using ultracold bosonic atoms in optical lattices
\cite{Experiments}, and this is what motivates this work. The
general guideline is that local or correlation functions will relax.
Most accessible would be a situation where only second moments would
have to be monitored, starting, e.g., with a checkerboard type
situation of alternating occupation numbers in a Mott state (cp.
Section~\ref{sec:moreGeneral} and Fig.~\ref{fig:secondmoments}),
prepared using two distinct optical lattices in a spatial dimension.
The idealized case of a vanishing interaction $U$ gives rise to a
guideline for what to expect in realistic settings and also to a
general principle: the local relaxation conjecture. It is our hope
that the present work fosters further experiments along these lines.

\begin{acknowledgments}
We acknowledge discussions with V.\ Buzek, A.\ Muramatsu, I.\
Peschel, M.M.\ Wolf,  M.B.\ Plenio, and B.\ Nachtergaele. This work
was supported by the DFG (SPP 1116), the EU (QAP), the QIP-IRC,
Microsoft Research, and the EURYI Award Scheme to JE and supported,
in part, by the Nuffield foundation for TJO. Note added: We have
become aware of Ref.\ \cite{Spohn} after finalizing this work. It
would be interesting to investigate whether, or to what extent, the
intuition of Ref.\ \cite{Spohn} for classical particles connected by
springs carries over to the quantum lattice case considered here.
\end{acknowledgments}

\end{document}